\documentclass[prl,showpacs,showkeys,nofootinbib,twocolumn]{revtex4}

\begin{document}

\noindent
{\bf Comment on ``Ramsey Fringes in a Bose-Einstein Condensate 
			between Atoms and Molecules''}
			
\hbox{$\ $}

In Ref.~\cite{Don02}, Donley et al.\ described an experiment at JILA
that demonstrated atom-molecule coherence in a Bose-Einstein condensate.
In a subsequent paper \cite{Holland}, Kokkelmans and Holland (K\&H) 
interpreted the results of Ref.~\cite{Don02} 
using a mean-field approximation to a resonance field theory  
involving the atom condensate $\phi_a({\bf r})$,
the molecular condensate $\phi_m({\bf r})$,
and the normal and anomalous densities 
$G_N({\bf r}_1,{\bf r}_2)$ and $G_A({\bf r}_1,{\bf r}_2)$.
Fig.~3 of K\&H shows $|\phi_a|^2$ and $|\phi_m|^2$
as functions of time for the pulse sequence in Fig.~2.
During the evolve time when the scattering length $a$ is constant,
$|\phi_a|^2$ oscillates around 0.80 with an amplitude of about 0.02,
while $|\phi_m|^2$ oscillates around 0.00325 with an amplitude 
of about 0.00030.  During the subsequent second pulse,
which takes $a$ first much closer to and then farther from 
the Feshbach resonance, $|\phi_m|^2$ decreases by a factor of 26 
and then increases by a factor of 118.
The interpretation of K\&H is that the molecular condensate density
is orders of magnitude smaller than the atom condensate density
during the evolve time and that it changes dramatically 
during the second pulse.  They conclude that
the observed oscillations involve a coherent flow 
of atoms between the atom condensate and the noncondensate atoms, 
with the molecular condensate playing only a minor role.

This interpretation is at best an awkward way 
of describing the physics.  It is important to distinguish between the 
diatomic molecule in the absence of hyperfine interactions
(the ``Feshbach molecule'') 
and in the presence of those interactions (the ``dimer'').  
The dimer is the energy eigenstate, and its binding energy
near the Feshbach resonance where $a$ diverges is $E_2 \approx \hbar^2/ma^2$.
The dimer can be expressed as a superposition of the Feshbach molecule 
and a 2-atom state:  it is mostly a Feshbach molecule 
far from the Feshbach resonance and mostly a 2-atom state 
near the Feshbach resonance.
If there is a molecular condensate, it will be a condensate of the dimers.  
Within the mean-field formalism of K\&H, 
the appropriate expression for the number density of atoms 
in the dimer condensate is $2 Z^{-1} |\phi_m|^2$,
where the renormalization constant $Z$ is the probability for the dimer 
to be a Feshbach molecule.

One can obtain an analytic expression for $Z$ 
in the resonance field theory, because it only requires solving 
a 2-body problem with contact interactions:
$$
Z^{-1} = 1 + {m^2 g_0^2 \over 16 \pi \hbar^3(m |E_2|)^{1/2}} 
	\left[ 1 - a_{\rm bg}(m |E_2|/\hbar^2)^{1/2} \right]^{-2},
$$
where $g_0$ is a coupling constant,
$a_{\rm bg}$ is the off-resonant scattering length,
and $E_2$ is the binding energy of the dimer.
The exact result for $E_2 = \hbar^2 \kappa^2/m$ 
is obtained by solving a cubic equation:
$$
(\kappa^2 + m \nu_0/\hbar^2) \left[ 1 - a_{\rm bg} \kappa \right]  
+ (m^2 g_0^2/8 \pi \hbar^4) \kappa = 0.
$$
The approximation of K\&H takes into account some many-body effects, 
but they are small enough in the JILA experiment
that the 2-body results for $Z^{-1}$ and $E_2$ are accurate.
Using the parameters in K\&H,
we find that $Z^{-1}$ is 18.3 during the evolve time, 
and during the second pulse it
increases to 1280 and then decreases to 3.7.
The resulting estimate $2 Z^{-1} |\phi_m|^2$ for the number density of atoms
in the dimer condensate during the evolve time oscillates around 0.119
with an amplitude of about 0.011.  The mean value
is about ${1 \over 7}$ the atom condensate density $|\phi_a|^2$,
and the amplitude of the oscillation is more than
${1 \over 2}$ that of $|\phi_a|^2$.
The changes in $Z^{-1}$ during the second pulse
cancel most of the dramatic changes in $|\phi_m|^2$,
so that $2 Z^{-1} |\phi_m|^2$ does not change dramatically.
In the pulse before the evolve time, a large fraction
of the atoms must have been transferred coherently from the atom condensate
to the dimer condensate by the strong resonant interaction
near the Feshbach resonance.    We believe a more accurate
treatment would reveal that the oscillations observed during the
evolve time involve the coherent flow of atoms between the
two condensates,  with noncondensate atoms playing only a minor role.

Our criticism of the interpretation of K\&H does not invalidate the agreement
between their results and the measurements of Ref.~\cite{Don02}.
K\&H found quantitative agreement between their results 
for $|\phi_a|^2$ after the second pulse
and the number density of the remnant BEC in the JILA experiment.
They also found qualitative agreement between their results
for $G_N({\bf r},{\bf r})$  after the second pulse
and the number density of burst atoms in the experiment.
In the formalism of K\&H, conservation of atoms implies 
$G_N({\bf r},{\bf r}) = n - |\phi_a|^2 - 2 |\phi_m|^2$,
where $n$ is the total number density of atoms.
After the second pulse, the system is relatively far from the 
Feshbach resonance, so $Z$ is close to 1 and  
$G_N({\bf r},{\bf r})$ is close to $n - |\phi_a|^2 - 2 Z^{-1} |\phi_m|^2$,
which is the correct expression 
for the number density of noncondensate atoms. 
Thus qualitative agreement between 
$G_N({\bf r},{\bf r})$ and the number density of burst atoms 
is to be expected.

\noindent
Eric Braaten,$^1$ H.-W. Hammer,$^2$ and M. Kusunoki$^1$\\
\phantom{Eric} $^1$Department of Physics, The Ohio State University, \\
\phantom{Eric$^1$} Columbus, OH\ 43210, USA\\
\phantom{Eric}
$^2$Helmholtz-Institut f{\"u}r Strahlen- und Kernphysik,\\
\phantom{Eric$^2$} Universit{\"a}t Bonn, 53115 Bonn, Germany

\noindent
Received:\\
DOI:

\end{document}